%% file: ms.tex
\documentclass{article}

\usepackage[utf8]{inputenc}

\usepackage{etoolbox}
\newtoggle{usingbibtex}
\toggletrue{usingbibtex}
\togglefalse{usingbibtex}

\iftoggle{usingbibtex}{
\usepackage[square,numbers]{natbib}
\bibliographystyle{unsrtnat}
}{
\usepackage[
backend=biber,
style=numeric,sorting=none, isbn=false, doi=true, eprint=true, url=false,
]{biblatex}
\AtEveryBibitem{%
  \clearlist{language}%
}
\addbibresource{library.bib}
}

\usepackage{times}
\usepackage{graphicx}
\usepackage{subfigure} 
\usepackage{bm}

\usepackage{algorithm}
\usepackage[noend]{algorithmic}

\usepackage{amsmath}
\usepackage{amssymb}
\usepackage{amsthm}
\usepackage{mathtools}
\usepackage{breqn}

\usepackage{siunitx}  

\usepackage{multirow}
\usepackage{tabularx,booktabs}

\usepackage[shortlabels]{enumitem}
\usepackage{fmtcount}
\usepackage{xspace}

\usepackage[section]{placeins}

\usepackage{xcolor}

\usepackage{hyperref}

\let\proglang=\textsf

\newcommand{\code}[1]{{\texttt {#1}}}
\let\proglang=\text

\newcommand{\figref}[1]{Fig.~\ref{fig:#1}}
\newcommand{\tblref}[1]{Table~\ref{tab:#1}}
\newcommand{\sref}[1]{Sect.~\ref{#1}}
\newcommand{\stepref}[1]{~\ref{#1}}
\newcommand{\eqnref}[1]{Eq.~\eqref{eq:#1}}

\newcommand{\formularef}[1]{Formula~\ref{eq:#1}}

\newcommand{\algref}[1]{Algorithm~\ref{alg:#1}}

\newcommand*{\Ie}{I.e.\@\xspace}
\newcommand*{\Eg}{e.g.\@\xspace}

\newcommand{\doi}[1]{\href{http://dx.doi.org/#1}{\normalfont\texttt{\@doi{#1}}}}

\newcommand{\Prob}{\ensuremath{\mathsf{P}}}
\newcommand{\CDF}{\ensuremath{\mathsf{CDF}}}
\newcommand{\SF}{\ensuremath{\mathsf{SF}}}
\newcommand{\PDF}{\ensuremath{\mathsf{PDF}}}
\newcommand{\PPF}{\ensuremath{\mathsf{PPF}}}
\newcommand{\ISF}{\ensuremath{\mathsf{ISF}}}
\def\SciPy{SciPy }

\newcommand{\expnum}[2]{\num[group-digits = false, output-exponent-marker = \ensuremath{\mathrm{E}}]{#1e#2}}

\newcommand{\kolmogorov}{\mathop{\mathrm{kolmogorov}}}
\newcommand{\kolmogi}{\mathop{\mathrm{kolmogi}}}
\newcommand{\kolmogorovcode}{\code{kolmogorov}}
\newcommand{\kolmogicode}{\code{kolmogi}}

\newtheorem{pvmalgorithm}{Algorithm}

\newcommand{\keywords}[1]{\par\addvspace\baselineskip
\noindent\keywordname\enspace\ignorespaces#1}
\ifx\mycmd\keywordname
\def\keywordname{{\bf Keywords:}}
\else
\fi

    \title{Computing the Cumulative Distribution Function and Quantiles of the limit of the Two-sided Kolmogorov-Smirnov Statistic}

    \author{Paul van Mulbregt\\
    pvanmulbregt@alum.mit.edu
    }

\begin{document} 

    \maketitle
    \input{abstract}

    \input{introduction}

    \input{ks_review}
    \input{kolmog_analysis}

    \input{kolmog_prop}
    \input{kolmogi_analysis}

    \input{kolmogi_prop}
    \input{kolmog_results}

    \section{Summary}

In some parts of its domain, the \CDF/\SF \/ is a sum of many relevant terms.
Using an alternate formula, based on Jacobi theta functions, reduces the number of relevant terms to no more than 4.

Using Newton-Raphson successfully to compute the \ISF \/ requires a good initial estimate, and a good approximation to the derivative.  
The contributions of the higher order terms in the Taylor series expansions for the quantiles can make the root-finding a little problematic.  
We showed how to generate a narrow interval enclosing the root, a good starting value for the iterations, and a way to calculate the derivative with little additional work, so that many fewer N-R iterations are required and the computed values have smaller errors.

\iftoggle{usingbibtex}{
\bibliographystyle{unsrtnat}
\bibliography{library}}
{
\sloppy
\printbibliography
}

\end{document}

%% file: abstract.tex
 \begin{abstract}
The cumulative distribution and quantile functions for the 
two-sided one sample Kolmogorov-Smirnov probability distributions 
are used for goodness-of-fit testing. 
The \CDF\/ is notoriously difficult to explicitly describe and to compute, and for large sample size use of the limiting distribution is an attractive alternative, with its lower computational requirements.
No closed form solution for the computation of the quantiles is known.
Computing the quantile function by a numeric root-finder for any specific probability may require multiple evaluations of both the \CDF\/ and its derivative.  Approximations to both the \CDF\/ and its derivative can be used to reduce the computational demands.
We show that the approximations in use inside the open source SciPy python software result in increased computation, not just reduced accuracy, and cause convergence failures in the root-finding.
Then we provide alternate algorithms which restore accuracy and efficiency across the whole domain.
\keywords{Two-sided Kolmogorov-Smirnov, probability, computation, approximations}
\end{abstract}

%% file: introduction.tex
\section{Introduction}

The Kolmogorov-Smirnov statistics $D_n$, $D_n^+$, $D_n^-$ are statistics that can be used as a measure of the goodness-of-fit between a sample of size $n$ and a target probability distribution.  Computation of the exact probability distribution for these statistics is a little complicated, but Kolmogorov and Smirnov showed that they had a certain limiting behaviour as $n \rightarrow \infty$.

To be used as part of a statistical test, either the value of the Survival Function (\SF) (or its complement the \CDF) 
needs to computable for a given value of $D_n$/$D_n^+$, or values need to be 
known corresponding to the desired critical probabilities (\Eg $p=0.1, 0.01, \ldots$).  
The quantile functions associated with these distributions can be used to generate a table of critical values, but they can also be used to generate random variates for the distribution, and have also found applications to rescaling of the axes in some types of graphing applications.

For the one-sided $D_n^+$, a formula is known which can be used to compute the \SF\@. But a simple formula for the two-sided $D_n$ is unknown.
The quantile functions do not have a closed-form solution, hence need to be calculated either by interpolating some known values, or by a numerical root-finding approach.
It is here especially that approximations may be used to reduce the computational requirements.

The \proglang{Python} package \SciPy  (v0.19.1) \cite{SciPy} provides the
\code{scipy.stats.kstwobign} class for the distribution of the two-sided $ \lim_{n\rightarrow\infty} \sqrt{n}D_n$, which in turn make calls to the ``C" library \code{scipy.special}, to calculate both the \SF\/ and its inverse, the \ISF\@.

An analysis of the approximations used in the algorithms determined that these approximations were
only valid in a subset of the domain, resulting in loss of accuracy and/or increase in computation.
Several causes of root-finding failure are identified.
Computation which takes time proportional to $\frac{1}{x}$ is exposed.

We then provide alternative algorithms which have lower relative error as well as lower (and bounded) computation.
For the quantile functions, the number of Newton-Raphson iterations is reduced by a factor of 6, with all convergence failures removed.  The number of terms needed to evaluate the \CDF/\SF\/ for $\lim_{n\rightarrow\infty} \sqrt{n}D_n$
 is also reduced by a factor of 6, and the relative error of the results improved, often by orders of magnitude.

This paper is organized as follows.  \sref{sec:ks_review} provides a quick review of Kolmogorov-Statistics with special emphasis on the formulae needed for computation.
\sref{sec:kolmogorov} provides an analysis of the formulae for computing the \CDF/\SF\/ of the two-sided $\lim_{n\rightarrow\infty} \sqrt{n}D_n$.
\sref{sec:kolmog_prop}  then analyses the \SciPy  implementation, and provides an alternate recipe for computing the \CDF/\SF\@.  
This is followed by an analysis \sref{sec:kolmogi} and recipe \sref{sec:kolmogi_prop} for the quantile function.
\sref{sec:kolmog_results} provides numeric results showing the change in performance resulting from use of these algorithms, 
along with interpretation of results. 

The formulae for computing the \SF/\CDF/\PDF\/ have been available for quite some time.  The novelty in this work is the analysis of the \SciPy  implementation and the details of the recipes, especially for the quantile functions. 
Explicit formulae for brackets containing the root are provided which enable root-finding algorithms to proceed with many fewer iterations.

The ``C" code computing the \CDF\/ \& \SF\/ for this distribution was written quite some time ago, 
when computers had considerably slower clock speeds and sample sizes were considerably smaller than they are today.
To a user of the software, the answers may have seemed plausible for most real-world inputs.

%% file: ks_review.tex

\section{Review of Kolmogorov-Smirnov Statistics}
\label{sec:ks_review}

In 1933 Kolmogorov \cite{kolmogoroff1933, kolmogoroff1941} introduced the the empirical cumulative distribution function (ECDF) for a (real-valued) sample $\{Y_1, Y_2, \dots, Y_n\}$, each $Y_i$ having the same continuous distribution function $F(Y)$.
He then enquired how close this ECDF would be to $F(Y)$.   Formally he defined
\begin{align}
F_n(y) & = \frac{1}{n} \#\left\{i : Y_i <= y\right\} \\
D_n & = \sup_y \left|F_{n}(y)-F(y)\right|
\end{align}

After wondering whether $\Prob \{D_n \leq \epsilon\}$ tends to 1 as $n\rightarrow\infty$ for all $\epsilon$, he then answered affirmatively with the asymptotic result  \cite{kolmogoroff1933, kolmogoroff1941}
\begin{align}
\lim_{n\rightarrow\infty} \Prob \{D_n \leq x\, n^{-1/2}\} & = L(x) = 1 - 2 \sum_{k=1}^{\infty} (-1)^{k-1} e^{-2k^2x^2}
\end{align}
Kolmogorov's proof used methods of classical physics.  Feller \cite{feller1948, feller1950} provided a more accessible proof in English.

\begin{figure}[!htb]
\centering
\includegraphics[scale=0.25]{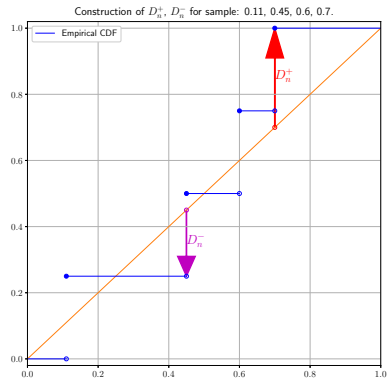}
 \caption{Construction of Kolmogorov-Smirnov statistics for $n=4$.}
\label{fig:KSconstruction}
\end{figure}

Smirnov \cite{zbMATH03107305, smirnov1948} instead used the one-sided values
$D_n^+ = \sup_y \left(F_{n}(y)-F(y)\right)$ and $D_n^- = \sup_y \left(F(y)-F_{n}(y)\right)$
 and showed that they also had a limiting form
\begin{align}
\label{eq:SmirnovAsymptote}
\lim_{n\rightarrow\infty} \Prob \{D_n^+ \leq x\, n^{-1/2}\} & = \lim_{n\rightarrow\infty} \Prob \{D_n^{-} \leq x\, n^{-1/2}\} = 1 - e^{-2x^2} \\
\shortintertext{Magg \& Dicaire\cite{Maag1971} gave a tightened asymptotic.  For a fixed $x$}
\label{eq:MaagAsymptote}
\Prob \{D_n^+ \leq x\} & \underset{{n\rightarrow \infty}}{\asymp}  1 - \exp\left({\frac{-(6nx+1)^2}{18n}}\right)
\end{align}
\figref{KSconstruction} illustrates the ECDF, and the construction of $D_n$, $D_n^+$ and $D_n^-$ for one sample.
The distributions of  $D_n^+$ and  $D_n^-$ are the same.  
The distributions of $D_n$ and $D_n^+$ are related, as $D_n = \max{D_n^-, D_n^+}$, and hence $\Prob \{D_n \geq x\} = 2*\Prob \{D_n^+ \geq x\} $ for all $x \ge 0.5$.

For the purpose of showing that the ECDF approaches $F(Y)$, these limit formulae are sufficient.
Later authors turned this around and used the $D_n$ statistic as a measure of ``goodness-of-fit" between the sample and $F(Y)$, for {\it any}\/ distribution function $F(Y)$.
It is clear that a large value for any of $D_n$, $D_n^+$ and $D_n^-$  may be indicative of a mismatch.
But a too-small value may also be cause for concern, as the fit may ``too good".
In order to use $D_n$ for this goodness-of-fit purpose, knowledge of the distribution of $D_n$ is needed.

Determining the exact distribution of the two-sided $D_n$ is non-trivial.
Birnbaum \cite{birnbaum1952} showed how to use Kolmogorov's recursion formulas to generate exact expressions for $\Prob(D_n \leq x)$, $n\leq6$.
Durbin \cite{durbin1968} provided a recursive formula to compute $\Prob(D_n \leq x)$ (implemented by Marsaglia, Tsang and Wang \cite{JSSv008i18}, made more efficient by Carvalho \cite{JSSv065c03}), which involved calculating a particular entry in a potentially large matrix raised to a high power.
Pomeranz \cite{Pomeranz:1974:AEC:361604.361628} provided another formulation which involved calculating a specific entry in a large-dimensional matrix.
Drew Glen \& Leemis \cite{Drew:2000:CCD:350528.350529} generated the collection of polynomial splines for $n<=30$.
Brown and Harvey \cite{JSSv019i06, JSSv026i02, JSSv026i03} implemented several algorithms in both rational arithmetic and arbitrary precision arithmetic.
Simard and L'Ecuyer \cite{JSSv039i11} analyzed all the known algorithms for numerical stability and sped.

For the one-sided statistics the situation is much cleaner.
An exact formula was discovered early-on \cite{zbMATH03107305, birnbaum1951, Miller1956}
 \begin{align}
 \label{eq:Sn1a}
 \Prob(D_n^+ \leq x) & = 1 - S_n(x)\\
 \shortintertext{where}
 \label{eq:Sn1}
 S_n(x) & =  x\sum_{j=0}^{\lfloor n(1-x)\rfloor} \binom{n}{j}  \left(x+\frac{j}{n}\right)^{j-1} \left(1-x-\frac{j}{n}\right)^{n-j}  \\
 \label{eq:Ajn}
 \end{align}
$S_n(x)$ is a sum of relatively simple $n$-th degree polynomials, forming a spline with knots at $0, \frac{1}{n}, \frac{2}{n}, \dots, 1$.
This has made computations involving $S_n(x)$ a somewhat easier task, though the simplicity can be a little misleading \cite{KSOneSided2018}.

%% file: kolmog_analysis.tex

\section{Computation of the Survival Function}
\label{sec:kolmogorov}

The \textrm{scipy.special} subpackage of the \textrm{Python}
\SciPy
  package provides two functions for computations of the limiting
$\sqrt{n}D_n$ distribution.
\code{kolmogorov(n, x)} computes the Survival Function $S_n(x)$ for $D_n^+$
and \code{kolmogi(n, p)} computes the Inverse Survival Function.
The source code for the computations is written in ``C", and are performed using the IEEE 754 64 bit double type 
(53 bits in the significand, and 11 bits in the exponent.)

The Survival Function is implemented directly as
\begin{align}
\label{eq:kolmogorov}
K(x) = 1-L(x) = \lim_{n\to\infty} \Prob\{\sqrt{n} D_n \geq x\}\, = \,2 \sum_{k=1}^{\infty} (-1)^{k-1}\, e^{-2k^2 x^2}
\end{align}


\subsection{Evaluating near x=0}

On the face of it, this appears to be a great series to be summing. 
It is an alternating series.  The exponent involves $-k^2$.  
And indeed for large $x$, the series converges quite quickly, needing only a few terms.  
But for small $x$, the situation is different.  For $x=0.01$ the first few terms in the series are:
$0.999800 - 0.999200 + 0.998201 - 0.996805 + 0.995012 - 0.992825 + 0.990247 - 0.987281 + 0.983930 - \ldots$.
After summing enough terms,  the result should be $0.5$ (actually $0.5 - \num[group-digits = false]{3.2d-5356}
$.)
Instead the sum is computed as $0.50000000000000044(=0.5+2^{-51})$
hence the calculated value of $K(x)$ is $1.0 + 2^{-50} > 1$.
This particular computed $\SF$ value has a {\em very}\/ small relative error, 
perhaps surprisingly small given that it has added up 400 terms and the condition number of the sum is 124,
However the value is outside the range of $K$,
and using it to compute the $\CDF$ has a rather large relative error.

Writing $q = e^{-2x^2}$, the sum is
\begin{align}
\label{eq:L_qsum}
K(x) & = 2 \sum_{k=1}^{\infty} (-1)^{k-1}\, q^{k^2} = q - q^4 + q^9 - q^{16}\ldots
\end{align}
The terms are certainly all smaller than 1, decreasing in size, with $\log({k\textrm{th}\, \text{term}}) = k^2 \log{(\text{1st term}})$.
However if the 1st term is too close to 1, then the terms decrease in magnitude {\em very}\/ slowly:  $q^{k^2} < \epsilon \iff k > \frac{\sqrt{\log{\frac{1}{\epsilon}}}} {\sqrt{2}x}$.
For $\epsilon=\num{e-16}$
say, a reasonable number for 64-bit floating point operations, $k > \frac{4}{x}$ to get to terms that are that small. 
In general, the number of terms needed is inversely proportional to $x$.  
For $x=0.01$, that requires summing over 400 mixed-sign terms, which provides many operations for rounding errors to accumulate, in addition to the loss of accuracy due to subtractive cancellation.
It is clear that a lot of precision is required to calculate an accurate value of $L(0.01)$.


\subsection{Combining adjacent terms}

One approach is to pair up the terms as per Monahan \cite{monahan1989}. This leads to
\begin{align}
\sum_{j=1}^{\infty} q^{(2j-1)^2}(1-q^{4j-1})  = q(1- q^3) + q^9(1-q^7) + \ldots
\end{align}
The example summation above would then become: $0.000600$ $+$ $0.001397$ $+$ 
$0.002187$ $+$ $0.002966$ $+$ $0.003732$ $+$ $0.004480$
$+$ $\ldots$
The terms are all positive, which provides some added stability, but still involves many terms.
This paired formulation actually requires more terms than the original \eqnref{L_qsum} in order for the terms to decrease enough in size,
but it has addressed the subtractive cancellation and now only suffers a precision issue.

Taking the combining a a step further, one can rewrite the sum as an infinite Horner method:
\begin{align}
K(x) & = {} 2 q(1-q^3(1-q^5(1-q^7(1-\ldots))))
\end{align}
Dropping the terms $q^{(2j+1)}(1-\dots)$ and beyond has an error less than $q^{(j+1)^2}$.
There is no difference in the number of terms required, but the powers of $q$ involved
are much lower, and truncating the computation provides an obviously non-negative answer.


\subsection{Alternate formulation via functional equation of Theta functions }

\begin{figure}[!htb]
\centering
\includegraphics[scale=0.5]{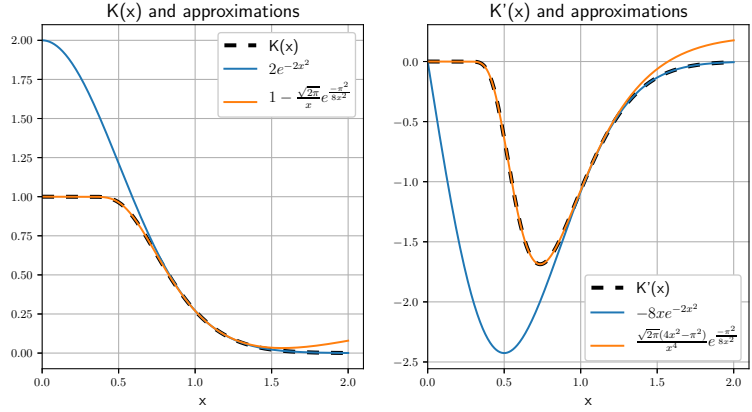}
\caption{$K(x)$ and two approximations; its derivative $K'(x)$ and two approximations.}
\label{fig:K_and_Kprime}
\end{figure}

The Kolmogorov probability can be expressed as a (Jacobian) theta function \cite{feller1948}:
\begin{align}
\Prob \{\sqrt{n} D_n \leq x\} \, & =\, 1 - 2 \sum_{k=1}^{\infty} (-1)^{k-1}\, e^{-2k^2 x^2} \,=\, \theta\left(z=\frac{1}{2}; \,\tau=\frac{2ix^2}{\pi}\right)
\shortintertext{where}
\theta\left(z; \,\tau\right) & = \sum_{k\in\mathbb{Z}} e^{\pi i k^2\tau + 2\pi i kz}  \text{\quad for $\tau \in \mathbb{H}$, the complex upper plane}
\end{align}
This theta function has a remarkable functional equation (\cite{whittaker1927course, mumford2006tata}), a simple form of which is:
\begin{align}
\theta\left( \frac{z}{\tau}; \, \frac{-1}{\tau} \right) &= e^{-\pi i /4} \sqrt{\tau} e^{\frac{\pi i z^2}{\tau}} \theta\left(z; \,\tau\right)
\shortintertext{After substituting values for $z, \tau$ and some simplification we arrive at}
\label{eq:L_small_x}
\Prob\{\sqrt{n} D_n \leq x\} \,=\, L(x) & =  \frac{\sqrt{2\pi}}{x} \sum_{k=1}^{\infty} e^{\frac{-(2k-1)^2 \pi^2}{8x^2}}\\
& =  \frac{\sqrt{2\pi}}{x} \sum_{\substack{n \in \mathbb{Z^+}\\n \; \text{odd}}}^{\infty} t^{n^2} \text{\quad where $t = e^{- \pi^2/8x^2}$}
\end{align}
This new formulation also contains a sum of some powers: $L(x) = t + t^9 + t^{25} + t^{49} + \dots$.
The difference in outcome is that for small $x$, the $t$ in this summation 
is {\em{much}}\/ smaller than the $q$ in \eqnref{L_qsum}, so these powers of $t$ become negligible after very few terms.  
For $x=0.01$ the first few terms are:
$\expnum{1.278}{-5358} + \expnum{9.105}{-48222}
+ \dots$
\Ie a single term is sufficient.
In general,
\begin{align}
t^{n^2} < \epsilon  \iff n > \sqrt{\frac{-8\log{\epsilon}}{\pi^2}} \, x
\end{align}
For $\epsilon=10^{-16}$ say,  $n > 6x$ will ensure small terms.  In fact, a single term will suffice for many $x$!
Because all these terms are positive, the sum is positive, and lies in the interval $[0, 1]$.
Combining terms in an infinite Horner method leads to an effective computation formulation
\begin{align}
\label{eq:L_small_xt}
L(x) &=  \frac{\sqrt{2\pi}}{x} t(1+t^8(1+t^{16}(1+t^{24}(1+\dots))))
\end{align}
\figref{K_and_Kprime} shows a plot of $K(x)$ and the two approximations arising from taking just the first term in the two series \eqnref{kolmogorov} and \eqnref{L_small_x}.
The approximations have different regions of quality, which fortunately overlap.

\subsection{Evaluation of the PDF}

The $\PDF$ can be evaluated by differentiating either \eqnref{kolmogorov} or \eqnref{L_small_x}
\begin{align}
PDF(x) & =  \frac{d\, \lim_{n\to\infty} \Prob\{\sqrt{n} D_n \geq x\}}{dx}\\
& = -K'(x) = L'(x)
\\ -K'(x) &= \,8x \sum_{k \in \mathbb{Z^+}} (-1)^{k-1}\, k^2 e^{-2k^2 x^2}
\\        & = 8x(q-4q^4+9q^9-16q^{16} \dots)
\\        & = 8xq(1-q^3(4-q^4(9-q^{7}(16-\dots))))
\label{eq:K_deriv}
\\ L'(x) & =  \frac{\sqrt{2\pi}}{4x^4}  {\sum_{\substack{n \in \mathbb{Z^+}\\n \; \text{odd}}} (-4x^2 + \pi^2 n^2) e^{\frac{-\pi^2n^2}{8x^2}}}
\\    & = \frac{-L(x)}{x}  + \frac{\sqrt{2}\pi^{5/2}}{4x^4}  (t + 9t^9 + 25t^{25} + 49t^{49} + \dots)\\
\label{eq:L_deriv}
\\    & =  \frac{-L(x)}{x} +  \frac{\sqrt{2}\pi^{5/2}}{4x^4}  t(1+t^8(9+t^{16}(25+t^{24}(49+\dots))))
\end{align}
Neither sum (\eqnref{K_deriv}, \eqnref{L_deriv}) has a high condition number when restricted to its appropriate regions.
The cost of evaluating $K'(x)$ is a little more the same cost of evaluating $K(x)$, and the 
cost of evaluating $L'(x)$ is a little more than the cost of evaluating $L(x)$.  Both sums are clearly positive.


\subsection{The SciPy implementation}

The SciPy implementation sums 
\eqnref{L_qsum}, until
$|\frac{q^{n^2}}{\text{partial sum}}| < \num{e-16}$,
with no recombination or use of alternate formulations.
For $x$ large enough, say $x > 0.75$ this works perfectly well.

\begin{itemize}
\item
As $x \downarrow 0$, the computation suffers from more and more accuracy loss due to subtractive cancellation.
For some values of $x$, the returned value is greater than 1.0.

\item
$K(x)$ is  {\em a priori}\/ a monotonically decreasing function of $x$, but the calculated values are far from monotonic.
As $x \downarrow 0$, the number of terms kept in the summation changes frequently (as it is inversely proportional to $x$.)
If the number of terms changes from even to odd, $K(x)$ jumps up.
Similarly if the number of terms becomes even, $K(x)$ jumps down.
Every time there is a switch, the monotonicity is lost (and also continuity!), so the results oscillate.
Root-finders appreciate the monotonicity as it makes that task easier.  
In this case a lack of it means that there may not be a solution to $K(x)-p=0$ inside an interval even though the value of the function at the endpoints have opposite signed values.
Another use of the $\ISF$ is to generate random variables from the distribution, given a random value in $[0, 1]$.  
For this it is is desirable to have the $\ISF$ be monotonic.  If the $\SF$ isn't monotonic then it's likely the $\ISF$ isn't monotonic either.

\item
SciPy doesn't provide a separate function to compute the $\PDF$.  Instead it numerically differentiates the $\CDF$ $1-K(x)$.  
This involves multiple evaluations of $K(x)$, and often returns negative values for $x \in [0, 0.2]$.
\end{itemize}

%% file: kolmog_prop.tex

\section[Algorithm for computing Kolmogorov SF, CDF and PDF]{Algorithm for computing Kolmogorov $\SF$, $\CDF$ and $\PDF$}
\label{sec:kolmog_prop} 
Here we propose an algorithm
to compute 
the $\CDF$/$\SF$ of the limit of the Kolmogorov two-sided statistic within a specified tolerance $\epsilon$, which addresses the issues discovered. 
It is an extension of Monahan's \SF\/ algorithm \cite{monahan1989}, to also cover the \PDF \/ and an arbitrary tolerance.

\setcounter{pvmalgorithm}{\value{algorithm}}
\begin{pvmalgorithm}[kolmogorov]
\label{kol_alg}
\label{kol_alg_api}
Compute the Kolmogorov  $\SF$, $\CDF$ and $\PDF$ for real $x$.
\end{pvmalgorithm}
\setcounter{algorithm}{\value{pvmalgorithm}}

\code{function [{SF}, {CDF}, {PDF}] = \kolmogorovcode(x:real, $\epsilon$:real)}
\noindent
\begin{enumerate}[start=1,label={{\bf Step \arabic*}},ref={Step \arabic*}]
\item  If $x \leq 0.82$ , set
\label{kol_alg_xlt82i}
\begin{align*}
        t & \leftarrow  {} \exp\left( \frac{-\pi^2}{8x^2} \right)
    \\   U & \leftarrow  {} \exp\left( \frac{-\pi^2}{x^2} \right)
    \\   R & \leftarrow  {} \left\lfloor{\sqrt{-2 \log(\epsilon)}*\frac{x}{\pi}+1} \right\rfloor 
    \\   S_R, D_R & \leftarrow  {} 1, (2R+1)^2\\
    \shortintertext{Then loop over $r \leftarrow R, R-1, \dots, 1$:}
    S_{r-1} & \leftarrow   1 + U^{r} * S_r
   \\    D_{r-1} & \leftarrow   (2r-1)^2 + U^{r} * D_r
    \shortintertext{Finally set}
    \code{CDF} & \leftarrow   \sqrt{2\pi}*t*S_0/x 
   \\    \code{SF} & \leftarrow  1 - \code{CDF} 
   \\  \code{PDF} & \leftarrow    \sqrt{2\pi}*t* \left( \pi^2D_0/(4x^2)  -  S_0\right)/x^2
\end{align*}

\item
If $x > 0.82$ , set
\label{kol_alg_xgt82i}
\begin{align*}
    q & \leftarrow   \exp\left({-2x^2} \right) \\
    R & \leftarrow   \left\lfloor{\frac{\sqrt{-2 \log(\epsilon)}}{3x}} \right\rfloor\\
    S_R, D_R & \leftarrow   1,  (R+1)^2\\
    \shortintertext{Then loop over $r \leftarrow R, R-1, \dots, 1$:}
    S_{r-1} & \leftarrow   {} 1 - q^{2r+1} * S_{r}\\
    D_{r-1} & \leftarrow  {} r^2 - q^{2r+1} * D_{r}\\
    \shortintertext{Finally set}
    \code{SF} & \leftarrow   2 * q * S_0\\
    \code{CDF}  & \leftarrow  1 - \code{SF}\\
    \code{PDF} & \leftarrow   8 * q * x * D_0
\end{align*}

\item
\label{kol_alg_returni}
Clip \code{SF}, \code{CDF} to lie in the interval $[0, 1]$ and \code{PDF} to lie in $[0, \infty)$.
Return \code{[{SF}, {CDF}, {PDF}]}.

\end{enumerate}


\subsection{Remarks}

Specifying the API to return both the $\CDF$ and $\SF$ probabilities enables
 the returned values to retain as much accuracy as had been computed.
It also enabled computing the $\CDF$ probabilities more accurately for values of $x$ close to 0.

\begin{itemize}
\item
If $x< 0.82$, \stepref{kol_alg_xlt82i} computes $p_{CDF}=L(x)$ using the series obtained from the functional equation for theta functions.
If $x>= 0.82$, \stepref{kol_alg_xgt82i} computes $p_{SF}=K(x)$.
This algorithm ensures that the values returned are between 0 and 1, with few items summed.
The cutoff of 0.82 is approximately the median of the distribution, so that the computation
 used will compute the smaller of $\{p_{SF}, p_{CDF}\}$,
  hence not incur loss if the complement needs to be returned.
For the smallest number of terms kept in the summation, the cutoff should be slightly higher, around 1.10 -- 1.15.
For the lowest error, the cutoff could be different for the $\CDF$ and the  $\SF$.

\item Calculation of the number of iterations required for the loop can be avoided by always looping the maximum number of times, which is  2 for \stepref{kol_alg_xlt82i} and 4 for \stepref{kol_alg_xgt82i}.  The loops can then be unrolled and the powers replaced with explicit squarings or cubings.

\end{itemize}


%% file: kolmogi_analysis.tex

\section{Evaluation of the Inverse Survival Function}
\label{sec:kolmogi}
 
Given a survivor probability $p_{SF}$, it is often desirable to know the $x$ that corresponds to it.  
There is no nice formula to invert $K(x)$, to go from a probability $p_{SF}$ back to a (scaled) difference $x$.

One way to numerically approximate the root of $K(x)= p_{SF} $ is to first solve for $q$:
\begin{align}
\label{eq:p_of_q}
\frac{p_{SF}}{2} =&\,  p = q - q^4 + q^9 - q^{16}\ldots
\shortintertext{and then set}
x & =  \sqrt{\frac{-\log(q)}{2}}
\end{align}
(or use this $x$ as the starting point for solving the original $K(x)=p_{SF}$).

Instead of addressing the full infinite sum, one can further approximate by truncating after $n$ terms.  Solve
\begin{align}
p = f_n(q) & = \sum_{k=1}^n (-1)^{k-1} q^{k^2} = q -q^4 + \dots + (-1)^{n-1}q^{n^2}
\end{align}
\begin{figure}[!htb]
\centering
\includegraphics[scale=0.5]{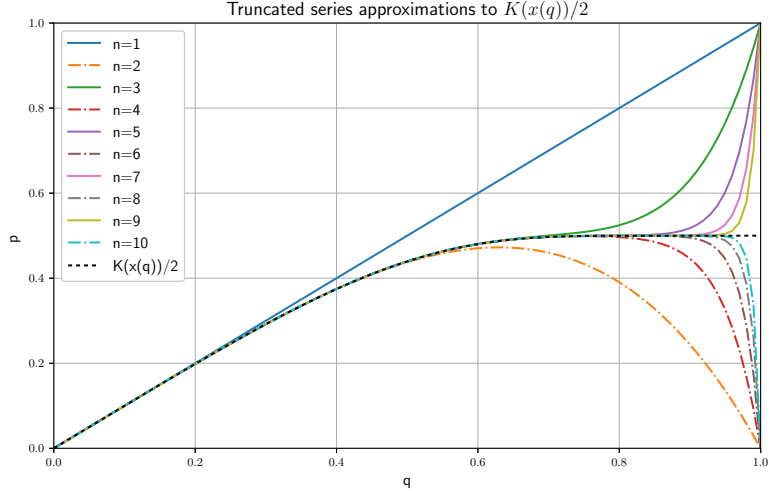}
\caption{Truncations to $K(x(q))/2$ after $n$ terms}
\label{fig:Ktruncate}
\end{figure}
\figref{Ktruncate} shows the graphs of the first few of these truncations.
All the curves start at the lower left $(q, p)=(0, 0)$, and head towards a height of $p=0.5$.
The $f_n$ for even $n$ top out a little below $p=0.5$, and then turn down to $(q,p)=(1, 0)$.
The $f_n$ for odd $n$ try to hug $p=0.5$ as long as they can, and then shoot up to $(q,p)=(1, 1)$.
Only the first truncation, $f_1$, deviates from the plot of $K(x(q))$ far from $p=0.5$.

When solving first for $q$ then setting $x$, a relative change of $\delta$ in $q$ results in a relative change
  in $x$ of $\frac{q \frac{dx}{dq}}{x}\delta =  \frac{1}{\log{q}}\delta=  \frac{-1}{4x^2}\delta$.
This can be used to guide the required tolerance when solving for $q$, 
 or provide an estimate of the error in $x=x(q)$ and hence 
 how much work needs to be done to polish it up.
%
%


\subsection[Use of root-solvers on f_n]{Use of root-solvers on $f_n$}

Inverting $f_1$ as a function is clearly trivial, but already inverting $f_2(q)=q-q^4$ requires some thought. 
 The maximum height of $f_2(q)$ is $3\cdot4^{-4/3}\approx 0.4724$ occurring at $q=4^{-1/3}$, 
 so not all the values of interest of $p\in [0, \frac{1}{2}]$ have a corresponding $q$.

One can use any polynomial root-finder to solve for $f_n(q)=p$.  
The non-zero roots of $f_n(z) = 0$ lie close to the unit circle in the complex plane, and for small $p$ the roots of $f_n(z) = p$ are just small perturbations of these.  
One or two steps of Laguerre iteration, or two or three steps of Newton-Raphson, provide a really good estimate $q_n$ solving $f_n(q_n) = p$.  
This in turn approximates the root of the original problem with small relative error (less than 1\%), at least until $p$ gets up to about 0.45.  
For these higher values of $p$, $n$ needs to be larger and larger in order to generate a good estimate for the original problem.


\subsection{Analytical inversion}
\label{kolmogi_analyticinversion}

Another approach is to treat the polynomial $f_n(z)$ as a function to $\mathbb{C}$ from the unit disk $D(0, 1)$ inside $\mathbb{C}$.
The function $f_n(z)$ is analytic, has non-zero derivative at $z=0$, and $f_n(z)=0$ hence
 there is a disk $D(0, r)$ and an analytic function $g(w)$, such that
$f(g(w)) = w$ for all $w\in D(0, r)$ (and $g(f(z)) = z$).

One way to find such a disk and $g$ is to treat the polynomial $f_n(q)$ as a formal power series, formally revert it and then analyze its region of convergence.
For $f_2(q) = q-q^4$, the formal power series reversion is
\begin{align}
\label{eq:f2_revert}
q=\sum_{k=0}^\infty {\binom{4k}{k}} \frac{p^{3k+1} }{3k+1} = p + p^{4} + 4p^{7}  + 22p^{10} + 140p^{13}+ 969p^{16}+\dots
\end{align}
This series converges for $|p| <3\cdot4^{-4/3} \approx 0.4724$.  
[If $p=0$, there are 4 solutions to $p=q-q^4=q(1-q^3)$, namely $q\in\{0, 1, \omega, \omega^2\}$, where $\omega=e^{-2\pi i/3}$ is a primitive cube root of 1.  
For small $p$, the series \eqnref{f2_revert} returns the root closest to 0. 
 As $p$ increases, this root increases towards 1, and the root which starts at 1 decreases towards 0.  
For $p=3\cdot4^{-4/3}$, these two roots meet at $q=4^{-1/3}$, and the power series reversion no longer converges.  
This also happens to be where the derivative $f_1'(q) = 1-4q^3=0$.  For $p>3*4^{-4/3}$ all roots are complex.]

\begin{table}[!htb]
\centering
\label{tab:fn_Inverse}
\begin{tabular}{ c | l }
$n$ & $f_n^{-1}(p)$ \\
\hline
1 & $p$
\\ 2 & $p  { \;}{\color{blue} +\; p^{4} + 4p^{7}  + 22p^{10} + 140p^{13}+ 969p^{16}+\dots}$
\\ 3 & $p + p^4 + 4p^7{\color{blue} {\mspace{4.0mu} -{\;} p^9}} + 22p^{10}
   {\color{blue} {\mspace{3.0mu}- \mspace{4.0mu} 13p^{12}}}    + 140p^{13}
   {\color{blue}\mspace{3.0mu} - \mspace{4.0mu} 136p^{15}}
   + 969p^{16} \dots$
  \\ 4 & $p + p^4 + 4p^7 - p^9 + 22p^{10}  - 13p^{12} + 140p^{13}  - 136p^{15} +  {\color{blue}970}p^{16} \dots$ \\
\hline
\end{tabular}
\caption{Formal power series inversion/reversion $f_n$ for $n=1, 2, 3, 4$.}
\end{table}

 \tblref{fn_Inverse} lists the first few terms for each of the first few  $f_n$, with changing coefficients highlighted.  The terms of $f_n^{-1}$ agree with those of $f_{n-1}^{-1}$ up to $p^{n^2-1}$.  Inverting the full $q$-series
 \begin{align}
 \label{eq:p_of_q_full}
 & p  = q - q^4+q^9-q^{16}+q^{25} - q^{36} + q^{49} - \dots \\
\shortintertext{leads to}
& \begin{aligned}
 \label{eq:p_reverse}
q  & =  {} p + p^{4} + 4p^{7} - p^{9} + 22p^{10} - 13p^{12} + 140p^{13} - 136p^{15} + 970p^{16}
\\  & + 9p^{17} - 1330p^{18} + 7104p^{19} + 231p^{20} - 12650p^{21} + 54096p^{22}
\\ &  + 3900p^{23} -  118780p^{24} + 423890p^{25} + 54810p^{26} - 1108380p^{27} + \dots
\end{aligned}
\end{align}

The coefficients  are all integers but they are not bounded, so the convergence properties of this expression are non-trivial to analyze.
The radius of convergence $r$ is no greater than $0.5$, hence the $n-$th coefficient must be about as big as $const\times2^n$ infinitely often.

While there is no universal formula to invert $2p = K(x)$, truncating this series \eqnref{p_reverse} and using for $p=\frac{p_{SF}}{2} \ll 0.5$ is reasonable.


\subsection{Bracketing the root}
\label{kolmogi_bracket}
Regardless of whether a good estimate of the root is available, an interval bracketing the root is useful as a guide/constraint on any numerical root-finders.  Starting from
\begin{align}
p_{SF} & = 2q (1-q^3+q^8 + \dots) \\
\shortintertext{rearranging as}
2q & = \frac{p_{SF}}{1-q^3+q^8 + \dots}\\
\shortintertext{and truncating the sum in the denominator, the following chain of inequalities emerge}
\label{eq:q_bounds_SF}
\frac{p_{SF}}{1} \leq \frac{p_{SF}}{1-q^3+q^8}\leq \dots  \leq & \quad 2q \quad \leq \dots \leq \frac{p_{SF}}{1-q^3+q^8-q^{15}} \leq \frac{p_{SF}}{1-q^3}
\end{align}
If $X_a$ is any fixed positive real number and $x>= X_a$, then $q=e^{-2x^2} < e^{-2X_a^2}$, and \eqnref{q_bounds_SF} provides bounds on $2q$ and hence on $x$.
Taking $X_a$ to be  median of the distribution, $X_a = K^{-1}(0.5) \approx 0.82757$, 
 with corresponding $q_a \approx 0.254 \approx e^{-1.37}$
\begin{align}
\sqrt{-0.5\log{\left(\frac{p_{SF}}{2}\right)}} \geq &\;\; x \;\;\geq \sqrt{-0.5 \log{\left(\frac{p_{SF}}{2 (1-e^{-4})} \right)}} & \text{\quad for $p_{SF} \leq 0.5$}
\end{align}
The upper bound is independent of any choice of $X_a$, but becomes less useful as $p_{SF}  \to 1$.


\subsection{Inverting for survivor probabilities close to 1}

For $p_{SF}$ close to 1, the approach taken in \eqnref{p_reverse} is not practical, but solving \eqnref{L_small_x} is.
\begin{align}
\label{eq:L_of_q}
p_{CDF} = L(x) & =  \frac{\sqrt{2\pi}}{x} \sum_{k=1}^{\infty} t^{(2k-1)^2}
\end{align}
Before applying Newton-Raphson to this we note that
\begin{align}
L'(x) & =  \frac{\sqrt{2\pi}}{4x^4}  {\sum_{\substack{n \in \mathbb{Z^+}\\n \; \text{odd}}} (-4x^2 + \pi^2 n^2) t^{n^2}}\\
L''(x) & = \frac{\sqrt{2\pi}}{16x^7} {\sum_{\substack{n \in \mathbb{Z^+}\\n \; \text{odd}}}  \left({32x^4 - 20\pi^2n^2x^2 + \pi^4 {n^4}}\right) t^{n^2}}
\end{align}
As $x \to 0$, $\frac{L''(x)}{2L'(x)} \sim \frac{\pi^2}{8x^3}$, which is unbounded.
Since the errors of each N-R step approximately follow $e_{n+1} \sim \frac{L''(x)}{2L'(x)} e_{n}^2$, 
 the initial error must already be very small in order to achieve rapid convergence.
Unfortunately there is no obvious good initial estimate for $x$ or $t$.  Rewrite \eqnref{L_of_q} as
\begin{align}
t & = \frac{x p_{CDF}}{\sqrt{2\pi}} \left({1 + t^8 +t^{24} + t^{48} + \dots }\right)^{-1}
\end{align}
Both $t$ and $L(x)$ are monotonically increasing functions of $x$.
If $X_a, X_b$ are any fixed positive real numbers and
 $ X_a \leq x \leq X_b (\iff L(X_a) \leq p_{CDF} \leq L(X_b))$, then
\begin{align}
 \frac{X_a p_{CDF}}{\sqrt{2\pi}} \; \frac{1}{1+t_b^8 + t_b^{24} \dots} \leq \; t \;
   & \leq  \frac{X_b p_{CDF}}{\sqrt{2\pi}} \;\frac{1}{1+t_a^8 + t_a^{24} \dots}
\end{align}
In particular, taking $X_b=1 (p_{b}\approx 0.73, t_b \approx 0.29)$, $X_a = 0.0406(t_a\approx \num{1e-325})$
\begin{equation}
\label{eq:ki_bigx_bound}
\left.
    \begin{aligned}
     \frac{0.04 * p_{CDF}}{\sqrt{2\pi}}  \leq t  & \leq  \frac{p_{CDF}}{\sqrt{2\pi}}  
     \\    \implies \frac{\pi}{\sqrt{-8 \log\left({\frac{0.04 * p_{CDF}} {\sqrt{2\pi}} }\right)}}  \leq \; x
      & \; \leq  \; \frac{\pi}{\sqrt{-8 \log\left({\frac{p_{CDF}} {\sqrt{2\pi}} }\right) } }
    \end{aligned}
\right\}
\quad \text{for $p_{CDF} \leq 0.73$}
\end{equation}
($X_a = 0.040596694\dots$ was used as $L(X_a)=2^{-1073}$,
so $L(x)$ is too small to be represented in 64 bits for any $x<X_a$.)
The upper bound is reasonable, but the lower bound is not so useful.

To find a better lower bound, drop all but the first summation term of $L(x)$ and solve
\begin{align}
p_{CDF} & =  \frac{\sqrt{2\pi}}{x} t =  \frac{\sqrt{2\pi}}{x} e^{\frac{-\pi^2}{8x^2}} \\
\shortintertext{which is equivalent to finding the fixed point $x_p$ of}
\label{eq:gp}
g_p(x) & \triangleq \frac{\pi}{\sqrt{-8\log{\left(\frac{p_{CDF} * x}{\sqrt{2\pi}}\right)}}}
\end{align}
The function is contractive around the fixed-point $x_p$ and the derivative there satisfies
\begin{align}
g_p'(x_p) & =  \frac{4x_p^2}{\pi^2} \ll 1
\end{align}
Starting with any value of $x$ and iterating with $g_p$ generates a monotonic sequence of values $\{x, g_p(x), g_p(g_p(x)), \dots\}$ converging to $x_p$. 
If $p_{CDF}$ is so small that only the first term in the summation contributes to the answer in machine precision, then $x_p$ is also the solution to $L(x)=p$.
[That occurs whenever
$t^8 < \epsilon \iff x < 0.523 \iff p_{CDF} < 0.0529$
 for the 64-bit floats.]
Starting with any upper bound for $L^{-1}(p_{CDF})$ (E.g. $x=1$), all the iterates will be upper bounds not only for $x_p$ but also $L^{-1}(p)$. [This is the same as the upper bound of \eqnref{ki_bigx_bound}.]
Starting with any of $x\in \{p_{CDF}, \sqrt{p_{CDF}}, 0.04\}$, the first few iterates will still be smaller than $L^{-1}(p_{CDF})$, so can be used for bracketing purposes.

Given the bracket, the bracket midpoint can be used as the starting point for methods such as N-R.
Though there may be some simple approximations for certain sub-intervals which lead to rapid convergence.  One example is:
\begin{align}
\label{eq:t_approx}
t & \approx 0.23530414p^2 + 0.2136641p - 0.00076411 \text{\qquad for $0.1 \le x \le 0.5$}
\end{align}


\subsection{The SciPy implementation}
\label{kolmogiSciPy}
The function $\kolmogi(p_{SF})$, in \SciPy's {\code{special}} sub-package returns an estimate of $K^{-1}(p_{SF})$.
It uses the Newton-Raphson method (without bracketing), approximates the derivative using just the
first term of \eqnref{kolmogorov} ($K'(x) \approx -8xe^{-2x^2}$), and halts whenever the relative change in the estimate is small enough ($|(x_{n+1}-x_n)/x_{n+1}| < \num{1e-10}$), or the number of iterations exceeds 500.  
The starting point for the N-R iterations is generated by using $f_1^{-1}$ to generate $q_0$, then $x_0= \sqrt{\frac{-\log(q_0)}{2}}$.

\begin{itemize}
\item This works well whenever $p_{SF}$ is small. 
In that situation, the first term in the summation dominates, so that $K(x) \approx 2q$.
Even though $x_0$ is greater than the desired root, the overshoot is small enough that the iterate stays in domain.

\item For $p_{SF}$ close to 1, the initial estimate $x_0$ is no longer close to the root, but this in itself is not a problem that the N-R couldn't overcome.
However the estimate used in place of $K'(x)$ is quite different from the actual value of $K'(x)$, as shown in \figref{K_and_Kprime}.
In particular, the true derivative is close to 0, while the approximation to the derivative may be orders of magnitude larger.
The result is that the adjustment made in each N-R step is {\em much}\/ too small.
This affects the convergence: not just the rate of convergence when the algorithm does converge to the actual root, but it gives the appearance of convergence even when still far from the root.
In particular, as $p_{SF}\to 1$, $\kolmogi(p_{SF})$ wouldn't return any number lower than 0.32, even though 0.18 should be achievable with 64 bits.

\item
The slow rate of convergence also quadratically affected the amount of computation needed.  
For $p_{SF}$ close to 1, $\kolmogi(p_{SF})$ required many iterations of N-R, each of which made a call to $\kolmogorov(x)$, and that in turn used many terms in its summation (as the number of terms $\propto \frac{1}{x}$).  
The net effect was that a single call to $\kolmogi(p_{SF})$ could generate 5000 calls to \code{exp}.

\end{itemize}

%% file: kolmogi_prop.tex


\section{Algorithm for Computing Kolmogorov Quantile}
\label{sec:kolmogi_prop}
Next we propose modifications to the existing algorithm
which will find $x$ such that $\kolmogorov(x) = p$ within the specified tolerances.

\setcounter{pvmalgorithm}{\value{algorithm}}
\begin{pvmalgorithm}[kolmogi]
\label{alg:kolmogi_alg}
\label{kolmogi_alg_api}
Compute the Kolmogorov $\ISF/\PPF$ for real $p_{SF}$, $p_{CDF}$.
\end{pvmalgorithm}
\setcounter{algorithm}{\value{pvmalgorithm}}
\code{function [$X$] = \kolmogicode(pSF:real, pCDF:real)}
\noindent
\begin{enumerate}[start=1,label={{\bf Step \arabic*}},ref={Step \arabic*}]
\item Immediately handle $p_{CDF}=0$, $p_{SF}=0$ as special cases, returning $X \leftarrow$ 0 or $\infty$ respectively.

\item Set
\begin{equation*}
\label{kolmogi_alg_deff}
f(x)= \begin{cases}
\code{kolmogorov(x).SF} - p_{SF} & \text{if $p_{SF}<= 0.5$}\\
p_{CDF} - \code{kolmogorov(x).CDF} & \text{otherwise}
\end{cases}
\end{equation*}

\item  If $p_{SF}\leq 0.5$, set
\label{kolmogi_alg_smallp}
\begin{align}
    P & \leftarrow  \frac{p_{SF}}{2}  \\
    \label{eq:q_A_B}
    [Q_A, \, Q_B] &\leftarrow  \left[ P * \frac{1}{1.0 - e^{-4}} , \, P \right] \\
    Q_0 & \leftarrow  P + P^{4} + 4P^{7}  - P^{9} + 22P^{10} - 13P^{12} + 140P^{13}   \\
    [A, \, B] &\leftarrow  \left[\sqrt{\frac{-\log{(Q_A)}}{2}}, \, \sqrt{\frac{-\log{(Q_B)}}{2}}\right] \\
    X_0 & \leftarrow  \sqrt{\frac{-\log{(Q_0)}}{2}}
\end{align}
Skip to \stepref{kolmogi_alg_nr}.

\item  If $p_{SF}> 0.5$, set
\label{kolmogi_alg_bigp}
\begin{align}
 [A_0, \, B_0] & \leftarrow   \left[ \max(\sqrt{p_{CDF}}, 0.04)  , \, 1 \right] \\
 [A, \, B ] & \leftarrow   \left[g_p(g_p(A_0)),  \,g_p(g_p(B_0))\right] \\
 X_0 &\leftarrow   \begin{cases}
0.2353p^2 + 0.2136p - 0.000764 & \text{if $p_{CDF} >= 0.1$}\\
 \frac{A + B}{2}& \text{otherwise}
 \end{cases}
\end{align}
Proceed to \stepref{kolmogi_alg_nr}.

\item  Perform iterations of bracketed N-R with function $f$, starting point $X_0$ and bracketing interval $[A, B]$, using the actual derivative of $f(x)$, until the desired tolerance is achieved, or the maximum number of iterations is exceeded.
 Set $X \leftarrow$ the final N-R iterate.  Return $X$.
\label{kolmogi_alg_nr}

\end{enumerate}


\subsection{Remarks}

\SciPy does contain multiple root-finders but we avoid
using them here.
The code for \kolmogorovcode{} is written in C as part of the {\code{cephes}} library in the \code{scipy.special} subpackage.
In order to enable an implementation of this K-S algorithm to remain contained within this subpackage,
we only use a bracketing Newton-Raphson root-finding algorithm, as this can be easily implemented.

Changing the API (\stepref{kolmogi_alg_api}) enables the code to use which ever probability allows the
greatest precision, which will usually be the smaller of the two.
It also enables computing $x$ more accurately for values of $p_{SF}$ close to 1.

\begin{itemize}
\item If $p_{SF}$ and $p_{CDF}$ are both non-zero, the root-finding will use a bracketed Newton-Raphson algorithm.

\item
In \stepref{kolmogi_alg_deff},  both expressions for $f(x)$ would compute the same answer if using infinite precision ---
any difference between them should be approximately the order of the machine epsilon.

\item For small $p_{SF}$ (which corresponds to $x\gtrapprox 0.82$) \stepref{kolmogi_alg_smallp} determines a good bracket and a good
initial estimate $X_0$.
The brackets come from \eqnref{q_bounds_SF}, the initial estimate from \eqnref{p_reverse}.

\item
In practice, it's been found that \eqnref{q_A_B} is a little tight for
some machine architectures/library implementations when dealing
with very small $p_{SF}$, and $Q_B$ should be a little smaller,
say $Q_B = \frac{p_{SF}}{2} * (1 - 256*\epsilon)$, where $\epsilon \approx  2^{-52}$ is the ``machine epsilon".
Similarly $Q_A$ should be a little larger.

\item For large $p_{SF}$ (which corresponds to $x\lessapprox 0.82$) \stepref{kolmogi_alg_bigp} first determines a good estimate and bracket for $q$,
and then $x$.
The brackets come from \eqnref{ki_bigx_bound}, the initial estimate from \eqnref{t_approx}.

\item  The N-R iterations require implementing a \code{kolmogorovp} function to calculate the derivative,
which can be done with the obvious minor modifications to \kolmogorovcode{}.
$f$ is $C^\infty$ so use of an order 1 method is justifiable.

 \end{itemize}

The main substance to the algorithm is the determination of a suitable bracket and initial starting point.
The choice of root-finder to complete the task is less important.

%% file: kolmog_results.tex


\section{Results}
\label{sec:kolmog_results}
\subsection{Kolmogorov}
\label{kolmogorov_results}

The methods compared are the SciPy v0.19 (``Baseline'') implementation, and an implementation of \algref{kolmogi_alg}, using
$x = 0, (0.001) 1.7$.

 \tblref{KolmogorovTable} shows some statistics for the number of summation terms used
 in the computation of the $\CDF$/$\SF$, (\formularef{L_qsum} or \formularef{L_small_xt})
 (The mean and std deviations are calculated over the values of $x$ which do not exceed 500 terms.
 The Failure column is the percentage of values of $x$ that exceeded 500 terms.
 The Tolerance column lists the percentage of values returned whose relative error exceeded \num{e-9}.)

 \begin{table}[!htb]
\centering
\begin{tabular}{r|rrr|rr}
\toprule
{} &  mean &  std &  max &  Failure &   Tolerance \\
\midrule
Baseline    &  12.4 &   31 &  481 &  0.4\% &  0.3\% \\
\algref{kolmogi_alg} &   2.2 &  0.9 &    4 &  0.0\% &  0.0\% \\
\bottomrule
\end{tabular}
\caption{Kolmogorov $\SF$: Summation Terms, Failure and Disagreement Rates}
\label{tab:KolmogorovTable}
\end{table}

\algref{kolmogi_alg} typically needs to evaluate just over 2 terms when
 computing the $\CDF$ or $\SF$ probabilities within a tolerance of $~\num{2.2e-16}$.
 (The maximum relative error in the computed value is actually higher than this,
 due to errors in $\log/\exp$, and roundoff in the multiplication/summation of the various terms.)
 This compares to an average of about 12 in the Baseline,
 which also uses a much higher tolerance of \num{e-10}.
The maximum number of iterations is also much reduced to about 4, with no failures to converge observed.
Most of the change in the number of iterations is due to using \formularef{L_small_xt}.


\subsection{Inverting Kolmogorov}
\label{kolmogi_results}

The methods compared are the SciPy (``Baseline'') implementation, and an implementation of \algref{kolmogi_alg}, using
$p = 0, (0.001) 1.0$.

 \tblref{KolmogiTable} shows statistics for the number of N-R iterations used in the computations of the $\ISF$/$\PPF$.
(The mean and std deviations are calculated over the values of $p$ which do not exceed 500 iterations.
The failure is the percentage of values of $p$ that exceed 500 iterations.
The tolerance column lists the percentage of values returned whose relative error exceeded \num{e-9}.)

 Typically 2-3 iterations are required for convergence within a tolerance of $2^{-52}\approx$ \num{2.2e-16} using \algref{kolmogi_alg},
 compared with 15 iterations (and a much higher tolerance of \num{e-10}) for the Baseline.
 The maximum is much reduced, with no failures to converge observed.
 (The maximum relative error in the computed value is potentially higher than $2^{-52}$, due to errors in $\log/\exp$,
 errors in computing $\kolmogorov(x)$ and roundoff in the multiplication/summation of the various terms.)

\begin{table}[!htb]
\centering
\begin{tabular}{r|rrr|rr}
\toprule
{} &  mean &   std &  max &  fail &   tol \\
\midrule
Baseline    &  15.5 &  27.3 &  379 &  0.2\% &  0.6\% \\
\algref{kolmogi_alg} &   2.5 &   0.9 &    4 &  0.0\% &  0.0\% \\
\bottomrule
\end{tabular}
\caption{Kolmogorov $\ISF$: Iteration Counts, Failure and Disagreement Rates}
\label{tab:KolmogiTable}
\end{table}

As noted in \sref{kolmogiSciPy}, the Baseline computation of $\kolmogi(p)$ increased quadratically as $p$ approached 1, as the number of iterations was essentially unbounded, and each iteration did an increasing amount of work.

\begin{itemize}
\item The starting point is close enough to the root that drastic overshoot below $x=0$ does not occur.  And it if did, the use of the bracket would prevent it.

\item Using the exact derivative extends the range of $\kolmogi(p_{SF})$ returnable values from $x\in [0.32, \infty)$ to $x\in [0.18, \infty)$ with 64 bit doubles.

\item Extending $\kolmogi(p_{SF})$ to $\kolmogi(p_{SF}, p_{CDF})$ allowed small $p_{CDF}$ values to be passed in exactly, rather than as $p_{SF} = 1-p_{CDF}$, and that extended the range of $\kolmogi(p_{SF}, p_{CDF})$ to $x\in [0.04, \infty)$ for
 64 bit doubles.

\item  If $p_{SF}$ is very close to 0, the number of N-R iterations required for solving $K(x)=p_{SF}$ is actually 0, as the initial estimate $\sqrt{(-0.5 \log(0.5 p_{SF}))}$ is {\em{a priori}} \/within tolerance.
For $p_{CDF}$ close to 0 ($p_{CDF}=\{2^{-n}: n=52, \dots 1022\}$), typically 2 iterations were required.

\item If the probabilities at smaller values of $x$ are needed, it becomes necessary to work with higher precision floats or work with log probabilities, which extend the domain beyond $x\in[0.04, \infty)$.
Since $g_p(x)$ only uses the log of $p_{CDF}$ and converges very quickly for very small $p_{CDF}$, iterating \eqnref{gp} is an efficient approach.

\item Other  root-finding algorithms could be used as $K(x)$ is well-behaved, when computed as above.
Given a tight starting bracket, Brent's method averaged 5.4 iterations,  Sidi's method \cite{Sidi2008a}
 (with $k=2$) and False Position with Illinois both averaged about 3.5 iterations.
An argument could be made that N-R requires two function evaluations per iteration,
for $K(x)$ and $K'(x)$, so that just counting iterations is underestimating the N-R work.
The additional work in simultaneously calculating the $\PDF$ with the $\CDF$/$\SF$ is small, so comparing the number of iterations between methods is reasonable.

\end{itemize}